\documentstyle[prl,aps,twocolumn,epsf]{revtex}

\def\be{\begin{equation}}
\def\ee{\end{equation}}
\def\bt{\begin{tabular}}
\def\et{\end{tabular}}
\def\lp{\left(}
\def\rp{\right)}
\def\ls{\left[}
\def\rs{\right]}

\def\bea{\begin{eqnarray}}
\def\eea{\end{eqnarray}}

\newcommand{\ts}{\textstyle}

\begin{document} \draft
\twocolumn[\hsize\textwidth\columnwidth\hsize\csname
@twocolumnfalse\endcsname


\title{
Vorticity affects the stability of neutron stars
}
\author{
Vahid Rezania$^{\dagger\ddagger}$ and Roy Maartens$^{\dagger}$
}
\address{~}
\address{
$\dagger$ Relativity and Cosmology Group, Division
of Mathematics and Statistics, \\ Portsmouth University,
Portsmouth~PO1~2EG, England
}
\address{
$\ddagger$ Institute for Advanced Studies in Basic Sciences,
Zanjan~45195, Iran
}
\address{~}
\date{\today}
\maketitle

\begin{abstract}
The spin rate $\Omega$ of neutron stars at a given temperature $T$
is constrained by the interplay between gravitational-radiation
instabilities and viscous damping. Navier-Stokes theory has been
used to calculate the viscous damping timescales and produce a
stability curve for $r$-modes in the $(\Omega,T)$ plane. In
Navier-Stokes theory, viscosity is independent of vorticity, but
kinetic theory predicts a coupling of vorticity to the shear
viscosity. We calculate this coupling and show that it can in
principle significantly modify the stability diagram at lower
temperatures. As a result, colder stars can remain stable at higher
spin rates.

\end{abstract}

\pacs{97.60.Jd, 04.40.Dg, 97.10.Sj }
]

It has recently been shown \cite{And,FM} that in rotating stars the
$r$-modes are unstable via a coupling to gravitational radiation.
This
instability can account for the spin-down of hot newly-born neutron
stars. Viscosity damps out the oscillations, and tends to stabilize
them in general. The timescales associated with this instability have
been calculated for uniform density and Newtonian polytropic stellar
models up to the lowest order of stellar spin rate $\Omega$
\cite{KS,Lin98}.

$r$-mode instability could be crucial in understanding the observed
spin rates of neutron stars at different temperatures, and it is
also a potentially observable source of gravitational radiation
\cite{O}. It is therefore important to improve our understanding of
the various factors that go into $r$-mode stability analysis. As
examples, the bulk viscous timescale has recently been calculated
to second order in $\Omega$ \cite{AKS,Lin99}, and relativistic
effects on $r$-modes have recently been considered \cite{LF}.
Further work remains to be done, both in refining the
thermo-hydrodynamical model, and in investigating relativistic
$r$-modes. In this Letter, we consider the possible effects of
vorticity on shear viscosity. These effects are predicted by
kinetic theory when the unperturbed equilibrium state is rotating.

In standard Navier-Stokes theory, the angular velocity of the fluid
has no effect on viscous stress or heat flux, which obey the
equations
\begin{equation}\label{ns} \Pi=-\zeta\Theta\,,~q_a=-\kappa\nabla_a
T\,,~ \pi_{ab}=-2\eta\sigma_{ab}\,, \end{equation} where $\Pi$ is the
bulk viscous stress and $\zeta$ is the bulk viscosity; $q_a$ is the
heat flux and $\kappa$ is the thermal conductivity; $\pi_{ab}$ is the
shear viscous stress and $\eta$ is the shear viscosity;
$\Theta=\nabla_av^a$ is the volume expansion rate of the fluid, $T$
is
the temperature, and $\sigma_{ab}=\nabla_{\langle a}v_{b\rangle}$ is
the rate of shear (where the angled brackets denote the symmetric
tracefree part). The fluid vorticity $\omega_a={1\over2}{\rm
curl}\,v_a$ does not enter these equations, even when the equilibrium
state is rotating.

On physical grounds, one might expect that rotational accelerations
can couple with gradients of momentum and temperature, so that there
could in principle be couplings of $\omega_a$ to $q_a$
and $\pi_{ab}$.
In the case of heat flux, qualitative particle dynamics indicates
\cite{MR} (p. 34) that this coupling does exist as a result of a
Coriolis effect, which is in some sense analogous to the Hall effect
in a conductor subject to a magnetic field. The Coriolis effect on
heat flux is confirmed by molecular dynamics simulations \cite{H}.
M\"uller \cite{M} and Israel \& Stewart \cite{IS} showed that the
Boltzmann equation predicts in general a coupling of vorticity to
heat
flux and shear viscous stress. The microscopic and self-consistent
kinetic approach is in contrast to the continuum view, where a
phenomenological principle of ``frame indifference" is invoked to
argue against any vorticity coupling. (See \cite{MR,H,JCL} for
further
discussion.)

Using the Grad moment method to approximate the hydrodynamic regime
via kinetic theory, the relations in Eq. (\ref{ns}) are modified to
\cite{IS} (Eq. (7.1))
\begin{eqnarray}
\Pi&=&-\zeta\left[\Theta+\beta_0\dot\Pi\right]\,,\label{1}\\
q_a&=&-\kappa\left[\nabla_a T+T\beta_1\left\{\dot{q}_a-\omega_{ab}q^b
\right\}\right]\,,\label{2}\\
\pi_{ab}&=&-2\eta\left[\sigma_{ab}+\beta_2\left\{\dot{\pi}_{\langle
ab\rangle} -2\omega^c{}_{\langle a}\pi_{b\rangle c}\right\}\right]\,,
\label{3} \end{eqnarray} where $\beta_A$ can be evaluated in terms of
collision integrals for specific gases, an overdot denotes the
comoving (Lagrangian) derivative, and the vorticity tensor is given
by
\[ \omega_{ab}=\nabla_{[a}v_{b]}=\varepsilon_{abc}\omega^c\,, \]
where square brackets on indices indicate the skew part.
Navier-Stokes theory is recovered from the M\"uller-Israel-Stewart
theory when $\beta_A=0$. However, kinetic theory gives $\beta_A$
values for simple gases which are definitely {\em not} zero.
Furthermore, if $\beta_A=0$, the equilibrium states are unstable
and dissipative signals can propagate at unbounded speed
\cite{IS,JCL}.

The $\beta_A$-corrections will be very small except if there are
either high frequency oscillations (pumping up the time-derivative
terms) or rapid rotation (pumping up the vorticity-coupling terms).
In the context of rapidly rotating neutron stars, we expect the
vorticity-dissipative couplings to dominate the time-derivative
terms; this expectation is borne out by calculations (see below).
The vorticity-dissipative couplings will be negligible if the
unperturbed equilibrium state is irrotational, i.e., if
$\omega_a=0$ in the background, so that the coupling terms become
second-order. However, for fast rotation, $\omega_a\neq0$ in the
background and the coupling terms make a first-order contribution
to dissipation. In the words of Israel \& Stewart \cite{IS}: \\
``these results will ultimately be of practical interest in
astrophysical and cosmological situations involving fast rotation,
strong gravitational fields or rapid fluctuations (neutron stars,
black hole accretion, early universe), although it will probably be
some time before the state of the art in these fields makes such
refinements necessary."\\ We believe that recent and ongoing
developments in rotating neutron star physics have reached the
stage where the M\"uller-Israel-Stewart theoretical corrections to
the Navier-Stokes equations need to be examined, and our results
indicate that the corrections could be important.

We follow the standard assumption \cite{CL} that the heat flux may be
neglected relative to viscous stresses in calculating damping
timescales. (Elsewhere we will discuss how the vorticity-heat flux
coupling could affect this standard assumption.) Then the vorticity
correction to Navier-Stokes theory reduces to the coupling term
$\omega^c{}_{\langle a}\pi_{b\rangle c}$. This term means that the
angular momentum of the star changes the shear viscosity timescale,
and we find (for axial $r$-modes) a correction proportional to
$T^{-r}\Omega^2$, where $r=9$ for a nonrelativistic fluid and $r=12$
for an ultrarelativistic fluid.

The evolution of dissipation energy contained in small fluctuations
is
given by \be \frac{d\tilde E}{dt}=-\int\ls\frac{|\delta\Pi|^2}
{\zeta}+
\frac{\delta\pi^{ab}\delta\pi_{ab}^*}{2\eta} \rs d^3x
-\lp\frac{d\tilde E}{dt}\rp_{\!\rm\sc g}\,,\label{ENERGY} \ee where
$(d\tilde E/dt)_{\rm\sc g}$ is the energy flux in gravitational
radiation, $\delta\Pi=\Pi-\bar\Pi$ and
$\delta\pi_{ab}=\pi_{ab}-\bar{\pi}_{ab}$, with an overbar denoting
background quantities. In this case, $\bar\Pi=0=\bar{\pi}_{ab}$.
The normal modes of the star are damped by dissipation, and the
damping rate can be determined by Eq. (\ref{ENERGY}). For a normal
mode with time dependence $e^{i\varpi t}$, the energy has time
dependence $\exp[-2{\rm Im}(\varpi) t]$. Then by Eq.
(\ref{ENERGY}), the characteristic damping time $\tau=1/{\rm
Im}(\varpi)$ of the fluid perturbation is given by \be
\frac{1}{\tau}=-{1\over2\tilde{E}}{d\tilde{E}\over dt}=
\frac{1}{\tau_{\rm\sc b}}+\frac{1}{\tau_{\rm\sc s}} +\frac{1}
{\tau_{\rm\sc g}}\,, \label{TOTALTIME} \ee where $\tau_{\rm\sc b}$,
$\tau_{\rm\sc s}$, and $\tau_{\rm\sc g}$ are the bulk viscous, shear
viscous, and gravitational radiation timescales respectively.

To evaluate the vorticity-corrected shear viscous timescale, we use
Eq. (\ref{3}) in Eqs. (\ref{ENERGY}) and (\ref{TOTALTIME}). To lowest
order
\[
\delta\pi_{ab}=-2\eta\left[\delta\sigma_{ab}-2i\varpi\eta\beta_2
\delta
\sigma_{ab} +4\eta\beta_2\delta\sigma^c_{\left<a\right.}
\omega_{\left. b\right>c}\right]\,, \] where $\omega_a$ is the
background vorticity (the background shear vanishes). Then
\begin{eqnarray*}
\delta\pi^{ab}\delta\pi^*_{ab}&=&4\eta^2\left\{\delta\sigma^{ab}
\delta\sigma^*_{ab}
+4\gamma^2\left[\varpi^2\delta\sigma^{ab}
\delta\sigma^*_{ab}
\right.\right.\\
&&\left.\left.{}+4\lp\delta\sigma^{ab}\delta\sigma^*_{ab}
\omega^c\omega_c
-\delta\sigma^{ca}\delta\sigma^*_{da}\omega_c\omega^d \rp
\right]\right\}\,,
\end{eqnarray*}
where $\gamma=\eta\beta_2$. The first term is the usual term in
Navier-Stokes theory, while the following terms are the
M\"uller-Israel-Stewart corrections. The $\varpi^2$ term arises
from $\dot{\pi}_{ab}$ in Eq. (\ref{3}), and is negligible relative
to the $\omega^2$ terms which arise from the $\omega^c{}_{\langle
a}\pi_{b\rangle c}$ term in Eq. (\ref{3}). The energy dissipation
rate through shear viscosity will be
\bea
\lp\!\frac{d\tilde E}{dt}\!\rp_{\!\!\rm\sc s}&=&
-2\!\int\! \eta\left\{
\delta\sigma^{ab}\delta\sigma^*_{ab}
-4\gamma^2\left[\varpi^2\delta\sigma^{ab}
\delta\sigma^*_{ab}\right.\right.\nonumber\\
&&\left.\left.{}+4\lp\delta\sigma^{ab}\delta\sigma^*_{ab}
\omega^c\omega_c
-\delta\sigma^{ca}\delta\sigma^*_{da}\omega_c\omega^d\rp
\right]\right\}d^3x.
\label{ENERGYSHEAR} \eea

In order to proceed further, we need expressions for the shear
viscosity $\eta$ and the coupling coefficient $\beta_2$. For the
various interactions, $\eta(\rho,T)$ is calculated in
\cite{FL76,FL79}, where it is shown that electron-electron
scattering is more important for shear viscosity than other
interactions. The expression for $\eta$ is given in \cite{CL},
in good
agreement with \cite{FL76,FL79}, as \be \eta=1.10\times
10^{16}\left(\!{ \rho\over 10^{14} {\rm g/cm}^3}\! \right)^{\!
9/4}\left(\!{10^9{\rm K}\over T}\!\right)^{\! 2}
\rm{g/cm\,s}\,.\label{ETA} \ee For a Maxwell-Boltzmann gas, the
coefficient $\beta_2$ is found in \cite{IS}, but we require the
expression for a degenerate Fermi gas. This has been found by
Olson \&
Hiscock \cite{OH} in the case of strong degeneracy:
\be
\beta_2={15\pi^2\hbar^3\over
m^4gc^5}\frac{(1+\nu)}{(\nu^2+2\nu)^{5/2}} +{\cal
O}\left[\left({kT\over mc^2\nu}\right)^{\!2}\right]\,, \ee where
$m$ is the particle mass, $g$ is the spin weight, and
$mc^2\nu/kT\gg1$. The dimensionless thermodynamic potential
$\nu=(\rho+p)/nm-mc^2s/kT-1$, where $s$ is the specific entropy, is
equal to the nonrelativistic chemical potential per particle
divided by the particle rest energy. For a strongly degenerate gas,
the nonrelativistic chemical potential is proportional to $T$, so
that
\[
\nu\approx\alpha \,{kT\over mc^2}\,,
\]
where $\alpha\gg1$ is a dimensionless constant measuring the degree
of degeneracy. The nonrelativistic regime is obtained for
$\nu\ll1$, while the ultrarelativistic case corresponds to $\nu\gg
1$.

For temperatures below $10^{10}$ K, neutrons in the neutron star
are nonrelativistic, while electrons are ultrarelativistic
\cite{FL76}. The nonrelativistic limit of $\beta_2$ is
\be
(\beta_2)_{{\rm\sc nr}}\approx 3.16 \times 10^{-5}(\alpha T)^{-5/2}~
\mbox{cm s}^{2}/\rm{g}\,,\label{BETA2NON} \ee and its
ultrarelativistic limit is \be (\beta_2)_{{\rm\sc ur}}\approx 6.45
\times 10^{15}(\alpha T)^{-4}~ \mbox{cm s}^{2}/\rm{g}\,.
\label{BETA2RE} \ee Using Eqs. (\ref{ETA}), (\ref{BETA2NON}) and
(\ref{BETA2RE}), we have \bea &&\gamma_{{\rm\sc nr}}\approx
{1.10\times 10^{-11}\over\alpha^{5/2}} \left(\!{ \rho\over 10^{14}
{\rm g/cm}^3}\!\right)^{\!9/4} \left(\!{10^9{\rm K}\over
T}\!\right)^{\!9/2} \rm{s}\,,\label{4}\\ &&\gamma_{{\rm\sc ur}}
\approx
{7.08\times10^{-5}\over\alpha^4} \left(\!{ \rho\over 10^{14} {\rm
g/cm}^3}\!\right)^{\!9/4} \left(\!{10^9{\rm K}\over T}\!\right)^{\!6}
\rm{s}\,. \label{5} \eea In the calculation, we used the same
relation
for $\eta$ in both cases, because in the high-density regime ($\rho >
10^{14}$g/ cm$^{3}$) for both electron-electron scattering and
electron-neutron scattering, $\eta$ is proportional to $T^{-2}$, with
nearly equal proportionality factor \cite{FL76}. For typical values
of
the temperature, $T=10^9$ K, and density, $\rho= 3\times 10^{14}$
g/cm$^{3}$, we find that $\gamma_{{\rm\sc ur}}\sim
\alpha^{-4}\times10^{-4}$ s, while $\gamma_{{\rm\sc
nr}}\sim\alpha^{-5/2}\times 10^{-10}$ s.

We assume that the background is a uniformly rotating star, so that
the equilibrium fluid velocity is $v^a=\Omega\varphi^a$, where
$\varphi^a$ is the rotational Killing vector field \cite{LF}. The
vorticity vector of the equilibrium state is \be
\vec\omega=\frac{\Omega}{2r}\ls\cot\vartheta\,, -1\,,0\rs\,.
\label{VORTICE1} \ee The $r$-modes of rotating barotropic Newtonian
stars have Eulerian velocity perturbations given by \cite{Lin98} \be
\delta\vec v= C R \Omega \lp\frac{r}{R}\rp^\ell\vec
{Y}^B_{\ell\ell}\exp(i \varpi t)\,,\label{VELOCITY} \ee where $C$
is an arbitrary constant, $R$ is the unperturbed stellar radius,
and $\varpi=2m\Omega/\ell(\ell+1)$. The magnetic-type vector
spherical
harmonics $\vec Y^B_{\ell m}$ are defined by \be
\vec{Y}^B_{lm}=\frac{r}{\sqrt{\ell(\ell+1)}}\vec\nabla\times\left[
 r\vec\nabla Y_{\ell m}(\vartheta,\varphi)\right]\,.
\ee
The shear of the perturbed star is given by
\begin{equation}
\delta\sigma_{ab}=\nabla_{\langle a}\delta
v_{b\rangle}\,.\label{6}
\end{equation}
Substituting Eqs. (\ref{VORTICE1})--(\ref{6}) into Eq.
(\ref{ENERGYSHEAR}), we find the shear viscosity timescale for
$\ell=m$:
\be
\frac{1}{\tau_{\rm s}}
\approx Q_\ell\left[(\ell-1)(2\ell+1)\int^R_0\eta r^{2\ell}\,dr
+\Omega^2 {\cal S}_\ell\right]\,,
\label{NEWSHEAR}
\ee
where
$Q_\ell^{-1}= \int_0^R\rho r^{2\ell+2}\,dr$.
The first term in brackets
is in agreement with the expression calculated in
\cite{Lin98}, and ${\cal S}_\ell$ is the correction term:
\begin{eqnarray} {\cal S}_\ell
&\approx&16{(\ell-1)(2\ell+1)\over(\ell+1)^{2}}U_0\nonumber\\
&&{}+\frac{\ell(\ell-2)![(2\ell-1)!!]^2}{(\ell+1)(2\ell-1)(2\ell)!}\,
\frac{\Gamma({\ts{1\over2}})} {\Gamma(\ell-{\ts{1\over2}})}
\times\nonumber\\
&&{}\times\left[ (2\ell^3-8\ell^2-3\ell-6)U_2
+12(\ell^3-\ell^2-\ell+1)U_3 \right.\nonumber\\
&&\left.{}+2(4\ell^4-\ell^3-9\ell^2+5\ell+1)U_4\right]\,,
\label{s}\end{eqnarray} where
$U_k(T)\equiv R^{k}\int^R_0 \gamma^2~\eta
r^{2\ell-k}\,dr$.

For the $\ell=2$ modes, Eqs. (\ref{NEWSHEAR})
and (\ref{s}) give
\begin{eqnarray}
\frac{1}{\tau_{\rm s}}&=& 5Q_2 \int^R_0\eta r^4 \,dr\nonumber\\
&&{}
+{\textstyle{1\over9}}Q_2\Omega^2 \left[80U_0+ 93U_2
+54U_3-42U_4\right]\,.
\label{NEWSHEAR1}
\end{eqnarray}
For comparison with previous calculations based on Navier-Stokes
viscosity (see, e.g., \cite{Lin99}), we use an $n=1$ polytrope
with mass $M=1.4 M_{\odot}$ and radius $R=12.57$ km to evaluate
the integrals in Eq. (\ref{NEWSHEAR1}). The bulk viscous and
gravitational radiation timescales are unaffected by the vorticity
correction, and we obtain \bea &&\frac{1}{\tau (\Omega,T)}
=\frac{1}{\tilde\tau_{\rm\sc g}}\lp\frac{\Omega}{\Omega_{\rm\sc
k}}\rp^6 +\frac{1}{\tilde\tau_{\rm\sc b}}\lp\frac{T}{10^9\rm
K}\rp^6\lp\frac{\Omega}{\Omega_{\rm\sc k}}\rp^2\nonumber\\ &&~~{}+
\frac{1}{\tilde\tau_{\rm\sc s}}\lp\frac{10^9\rm K}{T}\rp^2\ls 1+
q\alpha^{4-r}\lp\!\frac{10^9\rm K}{T}\!\rp^{\!r}\!
\lp\!\frac{\Omega}{\Omega_{\rm\sc k}}\!\rp^{\!2}\rs\,,
\label{RMODE} \eea where $\Omega_{\rm\sc k}=\sqrt{\pi G
\bar\rho}$, which is ${3\over2}$ times the Keplerian
(mass-shedding) frequency, and the vorticity correction factors
are
\begin{equation} q= \left\{\begin{array}{l}
    1.36 \times 10^{-23}\,,\\
    5.67\times 10^{-10}\,,
\end{array}\right.
~r=
\left\{\begin{array}{ll}
    9 & {\rm nonrel}\,,\\
    12 & {\rm ultrarel}\,.
\end{array}\right.
\label{7}
\end{equation}
The standard result (see, e.g., \cite{Lin99}) is regained for
$q=0$, with
\[
\tilde\tau_{\rm\sc g}=-3.26\,{\rm s}\,,~\tilde\tau_{\rm\sc
b}=2.01\times 10^{11} \,{\rm s}\,,~ \tilde\tau_{\rm\sc
s}=2.52\times 10^8\,{\rm s}\,.
\]
We note that the contribution from the $\dot{\pi}_{ab}$ term in Eq.
(\ref{3}) to the $q$-correction is less than 1\% of the
contribution from the $\omega^c{}_{\langle a}\pi_{b\rangle c}$
term.

Now we are able to determine from Eq. (\ref{RMODE}) the critical
angular velocity $\Omega_{\rm\sc c}$, defined by $1/\tau
(\Omega_{\rm\sc c},T)=0$, which governs stability of the star: if
$\Omega>\Omega_{\rm\sc c}$, then dissipative damping cannot
overcome the gravitational radiation-driven instability. In Fig.~1
we plot $\Omega_{\rm\sc c}/\Omega_{\rm\sc k}$ against temperature
$T$, showing how the vorticity-viscosity coupling affects the
standard result (see, e.g., \cite{Lin99}). Electrons are assumed to
dominate the shear viscosity, and they are ultrarelativistic over
the range of temperatures.

It is clear from Fig.~1 that the vorticity correction is only
appreciable at temperatures $T\lesssim 10^8$ K, but that for these
lower temperatures, the correction can be large, especially for
smaller $\alpha$. As the degree of degeneracy increases (i.e., with
increasing $\alpha$), the correction is confined to lower and lower
temperatures. The effect of the vorticity-viscosity coupling is to
increase the stable region, so that cooler stars can spin at higher
rates and remain stable. This may modify recent results
\cite{cold} which suggest that $r$-mode instability could stall
the spin-up of accreting neutron stars with $T\gtrsim 2\times10^5$
K; if the vorticity correction operates, then the stability region
is increased, so that spin-up could be more effective, especially
for lower degeneracy parameter $\alpha$.

Of course, our analysis is limited by the fact that we have
followed the standard assumption in viscous stability analysis and
ignored superfluid effects that will become important at lower
temperatures (see, e.g., \cite{lm}). Superfluid ``friction" effects
are thought to prevent $f$-mode instability, and these effects are
likely to be relevant also for $r$-modes. These effects may
strongly alter the vorticity correction effect, a subject which is
currently under investigation. In addition, we have used for our
$r$-mode calculations solutions that assume slow rotation. Thus the
$\Omega/\Omega_{\rm\sc k}\gtrsim 0.3$ part of Fig. 1 is an
extrapolation to high spin rates, in common with previous stability
diagrams. Recent calculations of $r$-modes for rapid rotation
\cite{li} should be used in future calculations of the vorticity
correction. Since $f$-modes are unstable at high spin rate, the
effect of the vorticity correction on these modes would also be
interesting to calculate.

In conclusion, we have shown that the coupling between vorticity
and shear viscous stress predicted by kinetic theory can in
principle have a significant effect on $r$-mode instability in
neutron stars. The M\"uller-Israel-Stewart correction of
Navier-Stokes theory predicts that colder stars can remain stable
at higher spin rates, so that accreting spin-up could be protected
from $r$-mode instability.

Finally, we remark that the vorticity correction to heat flux, as
well as the couplings between heat flux and viscous stress
predicted by kinetic theory \cite{M,IS} [but not shown in Eqs.
(\ref{1})--(\ref{3})], could lead to interesting modifications of
the standard stability curve. In particular, the coupling of bulk
viscous stress to heat flux could have an effect at {\em high}
temperatures \cite{rm}.

~\\ We are indebted to Nils Andersson and Sharon Morsink for
crucial comments, and we thank also Marco Bruni, Luis Herrera,
Mehdi Jahan-Miri, and Josep Triginer for helpful discussions. VR
was supported by a Royal Society grant.

~\\
\begin{figure}
\epsfxsize=3.4in \epsfysize=3.0in \epsffile{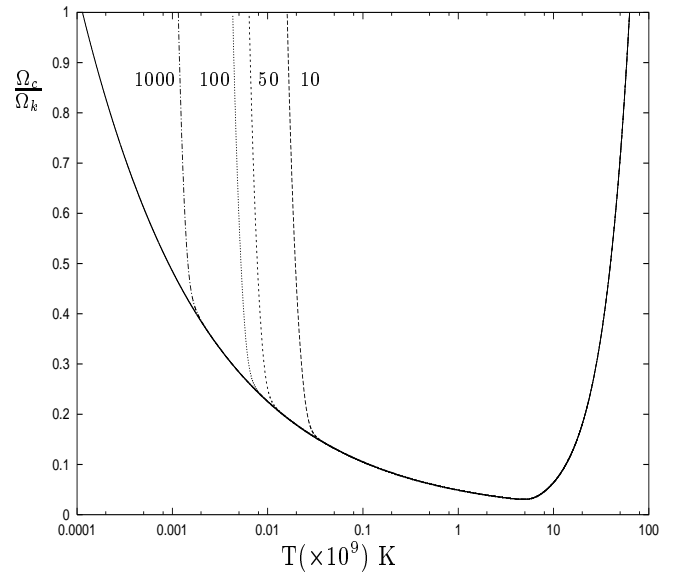} ~\\

\caption{Critical angular velocity versus temperature ($n=1$
polytrope with mass 1.4$M_\odot$ and radius 12.57 km). The
stability region is below the curves. The solid curve shows the
standard result, with no coupling of viscosity to vorticity.
Broken curves (labelled by the degeneracy parameter $\alpha$) show
how the instability region is reduced by the kinetic-theory
coupling of shear viscosity to vorticity, for an
ultra-relativistic degenerate Fermi fluid (electron-electron
viscosity). } \end{figure}


\end{document}